\documentclass{l4dc2027}

\title[ED-MPPI]{Explicit Dual Model Predictive Path Integral Control}
\usepackage{times}

\usepackage{mathtools}
\usepackage{array}
\usepackage{booktabs}
\usepackage{makecell}
\usepackage{multirow}
\usepackage{siunitx}
\usepackage{bbm}

\newcommand{\ms}[2]{$#1 \pm #2$}
\newlength{\srwidth}
\newcommand{\sr}[3]{\settowidth{\srwidth}{\small(00\,/\,000)}\addtolength{\srwidth}{2.2em}%
  \makebox[\srwidth][l]{\makebox[1.8em][r]{#1}~{\small(#2\,/\,#3)}}}
\newcommand{\best}[1]{\boldmath\bfseries #1}
\newcommand{\tbd}{--}

\coltauthor{
    \Name{Christopher Narr} \Email{c.narr@tum.de}\and
    \Name{Martin Buss} \Email{mb@tum.de}\\
    \addr Chair of Automatic Control Engineering, Technical University of Munich, Munich, Germany
}

\begin{document}

\maketitle

\begin{abstract}%
    In this work, we present a fully sampling-based explicit dual control algorithm that combines a particle filter and model predictive path integral control for partially observed nonlinear systems. The key idea is to roll out multiple sampled input sequences from all particles of the particle filter to obtain a belief-averaged control cost that induces caution. The same rollouts are used to sample imagined future measurements, which are drawn from the predictive distribution of the particle belief. This results in a Monte Carlo information term that gives a sample estimate of the expected information gain of each sampled input sequence and thereby induces probing. In contrast to existing explicit dual control schemes, the proposed scheme neither relies on gradients of the dynamics or measurement models nor requires a Gaussian belief. Furthermore, all computations can be parallelized across input sequences, particles, and imagined measurements, which makes the scheme amenable to real-time implementation on suitable hardware. We demonstrate the effectiveness of the proposed method in a simulation example of a planar terrain-aided navigation task.
\end{abstract}

\begin{keywords}%
    dual control, stochastic adaptive control, nonlinear predictive control 
\end{keywords}


\section{Introduction}
In many applications, a controller has to regulate a nonlinear system whose state or parameters cannot be measured directly and can only be inferred from noisy and possibly non-injective measurements, which makes learning about the system and controlling it inseparable tasks.
In this setting, the optimal stochastic output-feedback controller acts on the belief, i.e., the conditional probability distribution of the hidden variables given all past inputs and outputs~\citep{mesbah2018stochastic}.
Since the applied inputs determine which measurements are collected in the future, they also shape future beliefs, which is known as the \textit{dual effect}~\citep{feldbaum1960dual,bar1974dual}.
The optimal dual controller therefore exhibits two properties: it acts with \textit{caution} where the hidden variables are uncertain, and it \textit{probes} the system to reduce this uncertainty whenever the expected gain in future performance outweighs the resulting deviation from the control task~\citep{bar1976caution}.
In principle, the optimal policy can be computed by stochastic dynamic programming (SDP) in the belief space~\citep{feldbaum1960dual}, which, however, is tractable only for small problems, since the belief is in general infinite-dimensional.
Approximations are commonly classified into \textit{implicit} and \textit{explicit} dual control methods~\citep{mesbah2018stochastic}.

Implicit dual control methods approximate the SDP problem directly such that caution and probing emerge implicitly from their optimal solution.
For example, wide-sense dual control methods compute a series expansion of the cost-to-go around a nominal trajectory and propagate only the mean and variance of the belief inside the prediction~\citep{tse1973wide,klenske2016dual}.
Other implicit methods use techniques from approximate dynamic programming, such as scenario-based formulations~\citep{arcari2020dual,arcari2020approximate,hanssen2015scenario}, iterative dynamic programming~\citep{thompson2005stochastic}, or forward dynamic programming~\citep{bayard2010implicit}.
Online POMDP solvers with particle beliefs~\citep{silver2010monte,somani2013despot,sunberg2018online} can also be regarded as implicit dual controllers; they are gradient-free, handle multi-modal beliefs, and can include information rewards~\citep{araya2010pomdp,fischer2020information}, but their search trees grow quickly with the horizon for continuous actions and observations.
In general, the computational cost of implicit methods hinders their application to systems that require high control frequencies~\citep{mesbah2018stochastic}.

Explicit dual control methods circumvent the computational complexity by introducing a heuristic probing mechanism into the control formulation.
This can be achieved, e.g., by imposing persistent excitation constraints on the inputs~\citep{marafioti2014persistently}, by adding a cost term that penalizes predicted model or state uncertainty~\citep{heirung2015mpc,hovd2004interaction}, by maximizing a measure of the information matrix~\citep{rathousky2013mpc}, or by rewarding an information-theoretic measure of the expected learning~\citep{alpcan2015information}.
While computationally efficient, explicit methods often require differentiable models to construct the probing mechanism and are solved by gradient-based solvers, which struggle with highly nonlinear or non-smooth models and non-convex costs~\citep{rawlings2017model}.

Sampling-based approaches such as Model Predictive Path Integral (MPPI) control instead update a nominal input sequence by a cost-weighted average of sampled sequences~\citep{williams2018information}.
Since only forward evaluations of the dynamics are required, the rollouts can be parallelized, which has enabled real-time control of small-scale vehicles~\citep{williams2017information}, agile drones~\citep{zhai2026pa}, and legged robots~\citep{alvarez2025real}.
In output-feedback settings, MPPI is commonly combined with a filter whose point estimate is treated as exact, i.e., under certainty equivalence, while cautious MPPI variants~\citep{abraham2020model,das2026path} and particle-based output-feedback MPC~\citep{stahl2011pf,sehr2017particle} account for the uncertainty, but only learn passively, since their open-loop input sequences cannot anticipate future measurements~\citep{bar1974dual}.

To compute an explicit information term, a prediction of the future belief is required.
Dual control schemes commonly represent the belief and its predicted evolution by a Gaussian propagated through Kalman filter recursions~\citep{hovd2004interaction,arcari2020approximate,messerer2023dual}, which poorly approximates the multimodal beliefs that often arise from nonlinear dynamics and non-injective measurements.
Particle filters (PFs) represent such beliefs by weighted samples and only require forward evaluations of the dynamics and the measurement likelihood~\citep{chen2003bayesian}, which matches the computational structure of MPPI.
In dual control, PF beliefs have been combined with different mechanisms to predict the effect of future measurements on the belief:
\citet{flayac2017nonlinear,flayac2018dual} augment a PF-based controller with an information term derived from the Fisher information matrix, but require gradients of the measurement model;
\citet{chen2021dual} predict the PF posterior under hypothesized future measurements, but only for a finite set of admissible actions;
\citet{knaup2024active} use a PF within MPPI to estimate unknown parameters, but predict the future belief from the expected state under the current belief, similar to the maximum-likelihood observation assumption in belief-space planning~\citep{platt2010belief}, which induces probing only through a bias in the predicted belief rather than through a grounded information term.

In this work, we propose Explicit Dual MPPI (ED-MPPI), a fully sampling-based explicit dual controller that scores the information of continuous input sequences from the same particle rollouts that yield the control cost.
Our contributions are:
(i) a belief-averaged control cost and a Monte Carlo information term from shared particle rollouts, which add only one batched reweighting to cautious MPPI and require no gradients;
(ii) a proof that sampled imagined measurements preserve the predicted belief on average, unlike a single expected measurement, and that the information term is unbiased for the mutual information under the particle approximation;
(iii) a numerical comparison with certainty-equivalent, cautious, expected-measurement, and Fisher-information-based MPPI on a terrain-aided navigation task with smooth and piecewise-constant terrain.

\section{Problem Setup and Preliminaries}

\subsection{Problem Formulation}\label{sec:problem}
We consider discrete-time nonlinear systems of the form
\begin{equation}\label{eq:system_dynamics}
    x_{k+1} = f(x_k, u_k) + w_k, \qquad y_k = h(x_k) + v_k,
\end{equation}
with state $x_k \in \mathbb{R}^{n_x}$, input $u_k \in \mathbb{U} \subseteq \mathbb{R}^{n_u}$ from a closed and convex set $\mathbb{U}$, output $y_k \in \mathbb{R}^{n_y}$, and mutually independent i.i.d.\ noise $w_k \sim \mathcal{N}(0, Q)$, $Q \succeq 0$, and $v_k \sim \mathcal{N}(0, R)$, $R \succ 0$, and initial state $x_0 \sim p(x_0)$.
Crucially, $f$ and $h$ may be nonlinear and non-differentiable, and we only assume that forward evaluations are available (e.g., from a learned model or a simulator).
The controller only has access to the information vector $\mathcal{I}_k \coloneqq \{u_{0:k-1},\,y_{0:k}\}$ of all past inputs and outputs~\citep{bertsekas2017dynamic}.
The control task over the finite horizon $N \in \mathbb{N}$ is given by the cost function
\begin{equation}\label{eq:cost}
    J(x_k, u_{k:k+N-1}) \coloneqq \sum_{j=k}^{k+N-1} \ell_j(x_j, u_j) + V_f(x_{k+N}), 
\end{equation}
with a potentially time-varying stage cost $\ell_j: \mathbb{R}^{n_x} \times \mathbb{R}^{n_u} \to \mathbb{R}$ and terminal cost $V_f: \mathbb{R}^{n_x} \to \mathbb{R}$.
The states $x_{k+1:k+N}$ follow the dynamics~\eqref{eq:system_dynamics} from $x_k$ under the inputs $u_{k:k+N-1}$.
The cost $J$ also depends on the process noise $w_{k:k+N-1}$ through the states, which we omit in the notation for brevity.
State constraints $x_j \in \mathbb{X}$, with closed set $\mathbb{X} \subseteq \mathbb{R}^{n_x}$, are handled as penalty terms in the stage cost $\ell_j$, as commonly done in sampling-based MPC~\citep{williams2017information}.
Since $x_k$ is unknown and the states are subject to process noise, the stochastic output-feedback optimal control problem minimizes the expected cost over feedback policies, i.e.,
\begin{equation}\label{eq:ocp}
    \min_{\pi_k, \dots, \pi_{k+N-1}}\; \mathbb{E}\big[J(x_k, u_{k:k+N-1}) \,\big|\, \mathcal{I}_k\big] \quad \text{s.t.~\eqref{eq:system_dynamics}}, \quad u_j = \pi_j(\mathcal{I}_j) \in \mathbb{U}, \quad j = k, \dots, k+N-1.
\end{equation}
Problem~\eqref{eq:ocp} is intractable in general, since the policies act on information vectors of growing dimension and have to anticipate how every possible future measurement affects the subsequent inputs~\citep{mesbah2018stochastic}.
The first issue is addressed by replacing $\mathcal{I}_k$ by the belief $b_k \coloneqq p(x_k \mid \mathcal{I}_k)$, which is a sufficient statistic~\citep{bertsekas2017dynamic} and can be updated by Bayes' rule~\citep{chen2003bayesian}.
For nonlinear $f$ and $h$, however, this recursion has no closed-form solution, and the belief is in general non-Gaussian and possibly multimodal, such that it has to be approximated.

\subsection{Particle Filter}\label{sec:pf}
A PF approximates the belief $b_k$ by $N_p \in \mathbb{N}$ weighted particles $\mathcal{B}_k \coloneqq \{(x_k^{(l)}, \alpha_k^{(l)})\}_{l=1}^{N_p}$ with $x_k^{(l)} \in \mathbb{R}^{n_x}$, $\alpha_k^{(l)} \geq 0$, and $\sum_{l} \alpha_k^{(l)} = 1$, and can thus represent non-Gaussian and multimodal beliefs~\citep{chen2003bayesian}.
Using the transition density as proposal distribution, the time update of the filter propagates every particle through the dynamics~\eqref{eq:system_dynamics} with the applied input and sampled process noise, i.e.,
\begin{equation}\label{eq:pf_time_update}
    x_k^{(l)} = f\big(x_{k-1}^{(l)}, u_{k-1}\big) + w_{k-1}^{(l)}, \qquad w_{k-1}^{(l)} \sim \mathcal{N}(0, Q),
\end{equation}
and the measurement update reweights each particle by the likelihood of the new measurement,
\begin{equation}\label{eq:pf_meas_update}
    \alpha_k^{(l)} \propto \alpha_{k-1}^{(l)}\, \mathcal{N}\big(y_k \mid h(x_k^{(l)}),\, R\big).
\end{equation}
To counteract weight degeneracy, the particles are resampled by systematic resampling whenever the relative effective sample size $\text{rESS} \coloneqq 1 / (N_p \sum_{l} (\alpha_k^{(l)})^2)$ falls below a threshold, after which all weights are reset to $1/N_p$.
While the PF resolves the first issue, problem~\eqref{eq:ocp} still optimizes over policies, which can be approximated by optimizing over open-loop sequences instead.

\subsection{Open-Loop Approximations}\label{sec:approximations}
Tractable schemes replace the policies by an open-loop input sequence $u_{k:k+N-1} \in \mathbb{U}^N$, reoptimized at every time step in a receding-horizon fashion.
Certainty equivalence (CE) approximations additionally replace the belief by its mean $\hat{x}_k \coloneqq \mathbb{E}_{x_k \sim b_k}[x_k]$ and solve the deterministic CE problem $\min_{u_{k:k+N-1} \in \mathbb{U}^N} J(\hat{x}_k, u_{k:k+N-1})$ with $w_j = 0$, which yields neither caution nor probing.
Cautious approximations, such as ensemble MPPI (EMPPI)~\citep{abraham2020model}, optimize against the full belief by solving
\begin{equation}\label{eq:ocp_cautious}
        \min_{u_{k:k+N-1} \in \mathbb{U}^N}\; \mathbb{E}_{x_k \sim b_k, w_{k:k+N-1}}\big[J(x_k, u_{k:k+N-1})\big],
\end{equation}
which penalizes sequences that perform poorly for any likely state and thus yields caution.
However, for a fixed input sequence, the cost~\eqref{eq:cost} depends only on $x_k$ and the process noise, such that~\eqref{eq:ocp_cautious}, which is exactly problem~\eqref{eq:ocp} restricted to open-loop sequences, does not depend on $h$ and therefore does not probe~\citep{bar1974dual}.
Explicit dual schemes thus add an information term to the cost that rewards a predicted uncertainty reduction along the horizon~\citep{mesbah2018stochastic}.
For nonlinear and possibly non-differentiable $f$ and $h$, all of these costs can only be evaluated by forward simulation, such that gradient-free optimizers such as MPPI are a natural choice.

\subsection{Model Predictive Path Integral Control}\label{sec:mppi}
MPPI approximates the solution of a deterministic finite-horizon problem, such as the CE problem, from an initial state $x_k$ by a cost-weighted average of sampled input sequences~\citep{williams2017information}.
At each time step $k$, $K \in \mathbb{N}$ input sequences are sampled around a nominal input sequence $\bar{u} = \bar{u}_{k:k+N-1}$, i.e.,
\begin{equation}\label{eq:mppi_sampling}
    u^{(i)} = \Pi_{\mathbb{U}^N}\big(\bar{u} + \epsilon^{(i)}\big), \qquad i = 1, \dots, K,
\end{equation}
where $\epsilon^{(i)}$ is a zero-mean Gaussian noise sequence with $\epsilon_j^{(i)} \sim \mathcal{N}(0, \Sigma_\epsilon)$, $\Sigma_\epsilon \succ 0$, which is i.i.d.\ over $j$ in standard MPPI and may also be correlated over $j$ for smoother inputs, and $\Pi_{\mathbb{U}^N}$ denotes the Euclidean projection onto $\mathbb{U}^N$.
Each input sequence is rolled out from $x_k$ through the model dynamics~\eqref{eq:system_dynamics} with $w_j = 0$, which yields the rollout cost $S^{(i)} \coloneqq J\big(x_k, u^{(i)}\big)$ according to the cost~\eqref{eq:cost}.
The nominal sequence is then updated by the softmin-weighted average as
\begin{equation}\label{eq:mppi_update}
    \bar{u} \leftarrow \sum_{i=1}^{K} \omega^{(i)}\, u^{(i)}, \qquad \omega^{(i)} = \frac{\exp\big(-\tfrac{1}{\lambda}(S^{(i)} - S_{\min})\big)}{\sum_{i'=1}^{K} \exp\big(-\tfrac{1}{\lambda}(S^{(i')} - S_{\min})\big)},
\end{equation}
with a parameter $\lambda > 0$ and $S_{\min} \coloneqq \min_i S^{(i)}$ for numerical stability.
Since $\mathbb{U}$ is convex, the weighted average remains in $\mathbb{U}^{N}$.
The first input $\bar{u}_k$ is applied to the system, and the shifted sequence serves as the nominal input sequence at time step $k+1$.

\section{Explicit Dual Model Predictive Path Integral Control}
ED-MPPI scores every sampled input sequence by a belief-averaged control cost that induces caution (Section~\ref{sec:belief-cost}) and a Monte Carlo information term that induces probing (Section~\ref{sec:mc-info}), which are combined in Section~\ref{sec:ed-mppi}.

\subsection{Belief-Averaged Control Cost}\label{sec:belief-cost}
To account for the uncertainty in the belief $b_k$, every sampled input sequence $u^{(i)}$ is rolled out from every particle $l = 1, \dots, N_p$ of the PF as
\begin{equation}\label{eq:particle_rollout}
    x_{j+1}^{(i,l)} = f\big(x_j^{(i,l)}, u_j^{(i)}\big) + w_j^{(l)}, \qquad x_k^{(i,l)} = x_k^{(l)}, \qquad j = k, \dots, k+N-1,
\end{equation}
with sampled process noise $w_j^{(l)} \sim \mathcal{N}(0, Q)$.
Each step of~\eqref{eq:particle_rollout} is a PF time update~\eqref{eq:pf_time_update}, such that the propagated particle set $\{(x_j^{(i,l)}, \alpha_k^{(l)})\}_{l=1}^{N_p}$ approximates the open-loop predictive distribution $p(x_j \mid \mathcal{I}_k, u^{(i)}_{k:j-1})$.
The noise samples $w_j^{(l)}$ are drawn once per time step and shared by all input sequences as common random numbers, such that the cost differences between sequences result from the inputs only.
Evaluating the cost~\eqref{eq:cost} along each rollout~\eqref{eq:particle_rollout} yields $S^{(i,l)} \coloneqq J(x_{k}^{(l)}, u^{(i)})$, evaluated with the noise samples $w_{k:k+N-1}^{(l)}$, from which the belief-averaged control cost can be computed as
\begin{equation}\label{eq:ctrl_cost}
    S_{\text{ctrl}}^{(i)} \coloneqq \sum_{l=1}^{N_p} \alpha_k^{(l)}\, S^{(i,l)},
\end{equation}
which is a sample estimate of the expected cost in~\eqref{eq:ocp_cautious} and penalizes sequences that incur a high cost for any particle with non-negligible weight, which induces caution.

\subsection{Monte Carlo Information Term}\label{sec:mc-info}
To encourage input sequences whose future measurements reduce the uncertainty, we estimate their expected information gain by Monte Carlo sampling of imagined measurements, similar to particle-based information estimates in active sensing~\citep{ryan2010particle,hoffmann2010mobile} and Bayesian experimental design~\citep{foster2020unified,perez2025online}.
At each time step $k$, we draw $M \in \mathbb{N}$ imagined samples, each consisting of the index $l^\star_m \in \{1, \dots, N_p\}$ of an imagined true particle, with $\Pr(l^\star_m = l) = \alpha_k^{(l)}$, and a measurement noise sequence $v_j^{(m)} \sim \mathcal{N}(0, R)$, $j = k+1, \dots, k+N_I$, where $m = 1, \dots, M$ and $N_I \in \{1, \dots, N\}$ denotes the information horizon.
Note that $N_I$ must be at least the input-to-output delay of~\eqref{eq:system_dynamics}, since otherwise the inputs cannot affect the imagined measurements and the information term cannot induce probing.
Like the process noise, the imagined samples are shared by all input sequences.
With the outputs $\hat{y}_j^{(i,l)} \coloneqq h\big(x_j^{(i,l)}\big)$ of the rollouts~\eqref{eq:particle_rollout}, the imagined measurements of $u^{(i)}$ are the outputs of the imagined true particle plus noise, i.e.,
\begin{equation}\label{eq:imagined_meas}
    \tilde y_j^{(i,m)} = \hat{y}_j^{(i,l^\star_m)} + v_j^{(m)}, \qquad j = k+1, \dots, k+N_I.
\end{equation}
Based on the imagined measurements~\eqref{eq:imagined_meas}, we apply the measurement update~\eqref{eq:pf_meas_update} of the PF over the information horizon $N_I$, i.e., every particle is reweighted as
\begin{equation}\label{eq:imagined_weights}
    \tilde{\alpha}^{(i,l,m)} = \alpha_k^{(l)}\frac{\exp\big(\rho^{(i,l,m)}\big)}{Z^{(i,m)}}, \qquad Z^{(i,m)} \coloneqq \sum_{l'=1}^{N_p} \alpha_k^{(l')} \exp\big(\rho^{(i,l',m)}\big),
\end{equation}
where $\rho^{(i,l,m)} \coloneqq -\frac{1}{2} \sum_{j=k+1}^{k+N_I} \|\tilde y_j^{(i,m)} - \hat{y}_j^{(i,l)}\|_{R^{-1}}^2$ is the log-likelihood of particle $l$ up to an additive constant, which cancels in the normalization.
Since the rollouts~\eqref{eq:particle_rollout} are reused, the imagined update requires no additional evaluations of $f$, and since it omits resampling, the imagined weights remain defined on the same particles as $\alpha_k^{(l)}$, i.e.,~\eqref{eq:imagined_weights} is a sequential importance sampling update~\citep{chen2003bayesian}.
In the following, let $\tilde y \coloneqq \tilde y_{k+1:k+N_I}^{(i,m)}$ and let $L_l(\tilde y) \coloneqq \prod_{j=k+1}^{k+N_I} \mathcal{N}\big(\tilde y_j \mid \hat{y}_j^{(i,l)},\, R\big)$ denote the likelihood of particle $l$.
By~\eqref{eq:imagined_meas}, $\tilde y$ given $l^\star_m = l$ has the density $L_l$, such that $\tilde y$ follows the predictive density $p(\tilde y) = \sum_{l=1}^{N_p} \alpha_k^{(l)} L_l(\tilde y)$ of the propagated particle set, and the imagined weights~\eqref{eq:imagined_weights} are the exact posterior $\tilde{\alpha}^{(i,l,m)} = \alpha_k^{(l)} L_l(\tilde y) / p(\tilde y) = \Pr(l^\star_m = l \mid \tilde y)$.
Since the expected posterior of a Bayesian update equals the prior~\citep{williams1991probability}, any consistent prediction of future beliefs must preserve the current belief on average, which the imagined weights satisfy.
\begin{lemma}\label{prop:martingale}
    For a fixed input sequence $u^{(i)}$ and fixed rollouts~\eqref{eq:particle_rollout}, it holds $\mathbb{E}\big[\tilde{\alpha}^{(i,l,m)}\big] = \alpha_k^{(l)}$ for all $l = 1, \dots, N_p$, where the expectation is taken over $l^\star_m$ and $v_{k+1:k+N_I}^{(m)}$.
\end{lemma}
\textit{Proof:}
    The assertion follows from $\mathbb{E}_{\tilde y}\big[\tilde{\alpha}^{(i,l,m)}\big] = \int p(\tilde y)\, \alpha_k^{(l)} L_l(\tilde y) / p(\tilde y)\, \mathrm{d}\tilde y = \alpha_k^{(l)} \int L_l(\tilde y)\, \mathrm{d}\tilde y = \alpha_k^{(l)}$.
\hfill$\blacksquare$

In contrast, conditioning the belief only on the expected measurement $\bar{y}_j^{(i)} = \sum_{l} \alpha_k^{(l)} \hat{y}_j^{(i,l)}$, as in~\citet{knaup2024active}, yields deterministic weights that differ from $\alpha_k^{(l)}$ in general, such that the resulting predicted posterior belief is biased.

Each imagined sample $m$ yields one possible future belief $\{(x_{k+N_I}^{(i,l)}, \tilde{\alpha}^{(i,l,m)})\}_{l=1}^{N_p}$, and we quantify its information gain by the Kullback--Leibler (KL) divergence between the imagined and the predicted weights as
\begin{equation}\label{eq:imagined_gain}
    G^{(i,m)} \coloneqq \sum_{l=1}^{N_p} \tilde{\alpha}^{(i,l,m)} \log \frac{\tilde{\alpha}^{(i,l,m)}}{\alpha_k^{(l)}} = \sum_{l=1}^{N_p} \tilde{\alpha}^{(i,l,m)} \rho^{(i,l,m)} - \log Z^{(i,m)},
\end{equation}
with $\log Z^{(i,m)}$ computed via log-sum-exp.
The Monte Carlo information term then results from the average information gain over all imagined samples as
\begin{equation}\label{eq:info_term}
    S_{\text{info}}^{(i)} \coloneqq \frac{1}{M} \sum_{m=1}^{M} G^{(i,m)}.
\end{equation}

\begin{proposition}\label{prop:unbiased}
    Conditioned on the particle set $\mathcal{B}_k$, the process noise samples $w_{k:k+N_I-1}^{(l)}$, and the input sequence $u^{(i)}$, it holds $\mathbb{E}\big[S_{\text{info}}^{(i)}\big] = I\big(l^\star; \tilde y\big)$ for any $N_p$ and $M$, where $I(l^\star; \tilde y)$ denotes the mutual information between the particle index $l^\star \sim \alpha_k$ and the imagined measurements $\tilde y \sim L_{l^\star}$.
\end{proposition}
\textit{Proof:}
    By~\eqref{eq:imagined_weights} and~\eqref{eq:imagined_gain}, $G^{(i,m)}$ depends on $l^\star_m$ and $v_{k+1:k+N_I}^{(m)}$ only through $\tilde y \sim p(\tilde y)$.
    Since $\tilde{\alpha}^{(i,l,m)} = \Pr(l^\star_m = l \mid \tilde y)$ and $\alpha_k^{(l)} = \Pr(l^\star_m = l)$, $G^{(i,m)}$ is the KL divergence between the posterior and the prior of $l^\star_m$, whose expectation over $\tilde y$ is $I(l^\star; \tilde y)$~\citep{cover2006elements}.
    Since the $M$ imagined samples are identically distributed, the assertion follows from~\eqref{eq:info_term} by linearity of expectation.
\hfill$\blacksquare$

Since each particle carries its own noise samples, this is the information about the particle trajectory $x_{k:k+N_I}^{(i,l^\star)}$ rather than about $x_k$ alone.
With respect to the true belief, $l^\star_m$ is one of the particles in $Z^{(i,m)}$, such that, for equally weighted i.i.d.\ particles, $S_{\text{info}}^{(i)}$ equals the prior contrastive estimator of~\citet{foster2020unified}, i.e., a lower bound on the true mutual information in expectation that is tight only as $N_p \to \infty$ and upper-bounded by $\log N_p$.
Note that increasing $M$ reduces its variance but not its bias.
For small $R$ or long $N_I$, the imagined weights collapse onto few particles and $S_\text{info}^{(i)}$ saturates at this bound, which can be mitigated by likelihood tempering or a shorter $N_I$.

\subsection{ED-MPPI Algorithm}\label{sec:ed-mppi}
ED-MPPI replaces the rollout cost $S^{(i)}$ in the softmin weights~\eqref{eq:mppi_update} of the MPPI scheme by the cost
\begin{equation}\label{eq:total_cost}
    S_{\text{tot}}^{(i)} \coloneqq S_{\text{ctrl}}^{(i)} - \gamma\, S_{\text{info}}^{(i)},
\end{equation}
where the weight $\gamma \geq 0$ trades off the control task against uncertainty reduction.
In contrast to the cautious approximation~\eqref{eq:ocp_cautious}, the cost~\eqref{eq:total_cost} depends on $h$ through $S_{\text{info}}^{(i)}$ and thus rewards information acquisition.
As in other explicit dual schemes, a good trade-off between probing and control thus rests on the choice of $\gamma$.
Since $0 \le \mathbb{E}[S_\text{info}^{(i)}] \le \log N_p$, $\gamma \log N_p$ bounds the expected cost reduction that probing can achieve, which provides a scale for this choice.

\begin{algorithm2e}[!t]
    \caption{ED-MPPI}\label{alg:ed_mppi}
    \LinesNumbered
    \DontPrintSemicolon
    \KwIn{model $f$, $h$, covariances $Q$, $R$, costs $\ell_{k:k+N-1}$, $V_f$, horizons $N$, $N_I$, sample numbers $N_p$, $N_s \le N_p$, $K$, $M$, parameters $\lambda$, $\Sigma_\epsilon$, $\gamma$}
    Initialize particles $x_0^{(l)} \sim p(x_0)$ with weights $1/N_p$ and nominal sequence $\bar{u} \in \mathbb{U}^{N}$\;
    \For{$k = 0, 1, 2, \dots$}{
        Measure $y_k$, update particle weights via~\eqref{eq:pf_meas_update} (resample if required)\;
        If $N_s < N_p$, draw a subset of $N_s$ particles with weights $1/N_s$ via systematic resampling\;
        Draw process noise $\{w_{k:k+N-1}^{(l)}\}_{l=1}^{N_s}$ and imagined samples $\{l^\star_m, v_{k+1:k+N_I}^{(m)}\}_{m=1}^{M}$\;
        \For{$i = 1, \dots, K$ \textbf{in parallel}}{
            Sample input sequence $u^{(i)}$ via~\eqref{eq:mppi_sampling} and roll out input sequence $u^{(i)}$ over all particles via~\eqref{eq:particle_rollout}\;
            Compute $S_{\text{ctrl}}^{(i)}$ via~\eqref{eq:ctrl_cost}, $S_{\text{info}}^{(i)}$ via~\eqref{eq:imagined_meas}--\eqref{eq:info_term} (in parallel over $m$), and $S_{\text{tot}}^{(i)}$ via~\eqref{eq:total_cost}\;
        }
        Update $\bar{u}$ via~\eqref{eq:mppi_update} with $S_{\text{tot}}^{(i)}$ in place of $S^{(i)}$\;
        Apply $u_k = \bar{u}_k$ and propagate particles via~\eqref{eq:pf_time_update}\;
        Shift $\bar{u}$ by one step and repeat its last element\;
    }
\end{algorithm2e}

Algorithm~\ref{alg:ed_mppi} summarizes the overall scheme, whose computations are independent across input sequences, particles, and imagined samples, except for the sums over particles, and can thus be parallelized on a GPU.
To reduce the computational load, each input sequence can be rolled out only from a subset of $N_s \ll N_p$ particles with weights $1/N_s$, drawn once per time step by systematic resampling (line~4), such that $N_s$ replaces $N_p$ in~\eqref{eq:ctrl_cost}--\eqref{eq:info_term} and in the bound $\log N_p$.
However, while $S_{\text{ctrl}}^{(i)}$ remains unbiased, the variance of both terms increases.

\section{Simulation Example}
\subsection{Setup}
We evaluate the proposed method in a planar terrain-aided navigation task similar to~\citet{flayac2017nonlinear}, in which a robot with an ambiguous initial position has to pass a narrow gap in a wall.
The robot is modeled as a unicycle with dynamics
\begin{equation}
    p_{k+1} = p_k + \Delta T\, [\cos\theta_k, \sin\theta_k]^\top u_k^{v}, \qquad \theta_{k+1} = \theta_k + \Delta T\, u_k^{\omega},
\end{equation}
where the position $p_k = [p_k^{x}, p_k^{y}]^\top$ and the heading $\theta_k$ constitute the state $x_{k} = [p_k^\top, \theta_k]^\top$, and $\Delta T = \qty{0.1}{\second}$ is the sampling time.
The inputs are the forward velocity $u_{k}^{v} \in [-0.5, 1.0]\,\unit{\meter\per\second}$ and the angular velocity $u_{k}^{\omega} \in [-1.5, 1.5]\,\unit{\radian\per\second}$.
The state is additionally subject to process noise $w_{k} \sim \mathcal{N}(0,Q)$ with $Q = 10^{-5}\cdot\text{diag}(4,4,1)$.
Due to the small $Q$, the PF adds small Gaussian noise to the resampled particles (roughening)~\citep{chen2003bayesian}.
The robot only measures the terrain height $y_k = h(p_k) + v_k$ with $v_k \sim \mathcal{N}(0, R)$, $R = 2.5 \cdot 10^{-3}$, where $h(p) = 0$ behind the wall ($p^x \geq \qty{6}{\meter}$) and we consider two terrains in front of it:
the smooth \textsc{ramp} $h(p) = 0.05 \log\big(1 + \exp(10(p^y - 2))\big)$, which is flat around the center line ($p^y = 0$) and rises with slope $0.5$ for $p^y > \qty{2}{\meter}$, and the \textsc{step} $h(p) = 0.25$ for $p^y > \qty{2}{\meter}$ and zero otherwise, which is discontinuous at $p^y = \qty{2}{\meter}$ and has a zero gradient elsewhere.
The wall occupies $\qty{6}{\meter} \leq p^x \leq \qty{6.4}{\meter}$ except for a gap of width $\qty{1}{\meter}$ centered at $p^y = 0$.
The task is to reach $p_\text{goal} = [10, 0]^\top$ without colliding with the wall, encoded by the costs
\begin{equation}
    \ell(x, u) = d(p) + 0.1 \|u\|^2 + 200\, \mathbbm{1}_W(p), \qquad V_f(x) = 10\, d(p) + 200\, \mathbbm{1}_W(p),
\end{equation}
where $d(p)$ is the path length to $p_\text{goal}$ through the gap center $p_\text{gap} = [6.2, 0]^\top$, i.e., $d(p) = \|p - p_\text{gap}\| + \|p_\text{gap} - p_\text{goal}\|$ for $p^x < \qty{6.2}{\meter}$ and $\|p - p_\text{goal}\|$ otherwise, and $\mathbbm{1}_W$ is the indicator function of the wall inflated by $\qty{0.05}{\meter}$.
A run is successful if the robot passes the wall within $\qty{20}{\second}$, a collision if it hits the wall, and a timeout otherwise.
The initial belief is $b_0 = \tfrac{1}{2}\mathcal{N}(\mu^+, \Sigma_0) + \tfrac{1}{2}\mathcal{N}(\mu^-, \Sigma_0)$ with $\mu^\pm = [0, \pm 1, 0]^\top$ and $\Sigma_0 = \text{diag}(0.1^2, 0.1^2, 0.03^2)$, and the true initial state lies in the upper mode in half of the runs and in the lower mode in the other half.
Neither mode is aligned with the gap, and the belief mean lies on the flat center line, such that the robot can only pass safely after resolving the ambiguity by a detour towards the ramp or step.

We compare the proposed ED-MPPI to four baselines, which use the same PF and MPPI optimizer and differ only in the rollout cost that enters the softmin weights~\eqref{eq:mppi_update}.
CE-MPPI rolls out each input sequence only from the PF mean $\hat{x}_k = \sum_{l=1}^{N_p} \alpha_k^{(l)} x_k^{(l)}$ with $w_j = 0$, and thus approximately solves the CE problem.
Cautious-MPPI uses the belief-averaged control cost~\eqref{eq:ctrl_cost} as rollout cost, i.e., it corresponds to ED-MPPI with $\gamma = 0$, and approximately solves the cautious problem~\eqref{eq:ocp_cautious}; it is similar to EMPPI~\citep{abraham2020model} with state particles in place of parameter samples.
DMPPI~\citep{knaup2024active} averages the stage costs along each rollout over a predicted belief whose weights are updated with the expected measurement $\bar{y}_j^{(i)}$, $j = k+1, \dots, k+N-1$, applied to state instead of parameter particles.
Fisher-MPPI augments Cautious-MPPI with the gradient-based information term of~\citet{flayac2017nonlinear}, which penalizes the trace of the inverse Fisher information matrix of the position, summed over the information horizon.
Finally, Oracle-MPPI applies CE-MPPI to the true state as a non-implementable reference.
All methods share the PF with $N_p = 2000$ particles and the MPPI parameters $N = 60$, $K = 256$, $\lambda = 1$, $N_s = 256$, and $\Sigma_\epsilon = \text{diag}(0.25, 1)$ with a correlation of $0.9$ between consecutive steps.
ED-MPPI uses $M = 16$, $N_I = 40$, and $\gamma = 20$, while Fisher-MPPI uses $N_I = 40$ and $\beta = 10$ with batched analytic Jacobians.
All methods are evaluated over $N_{\text{MC}} = 100$ Monte Carlo simulations, implemented in Python with PyTorch and run on an NVIDIA GeForce RTX 5070 GPU. 

\begin{figure}[!t]
    \centering
    \includegraphics{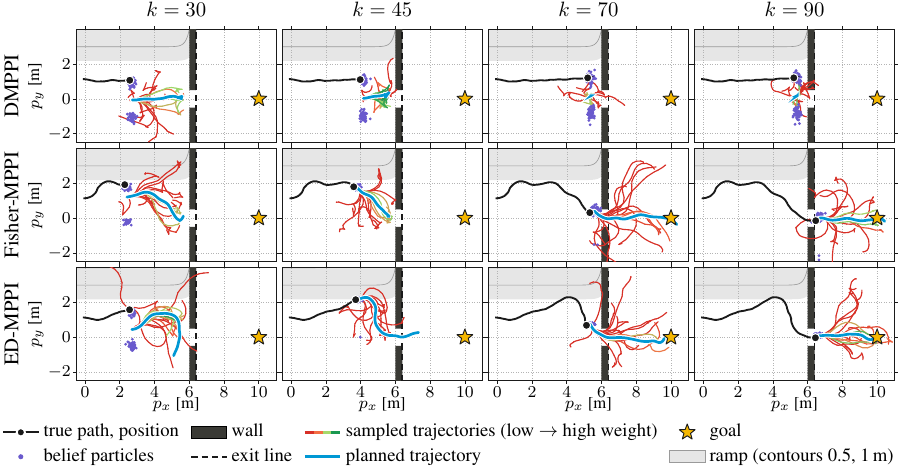}
    \caption{Closed-loop snapshots of DMPPI, Fisher-MPPI, and ED-MPPI at selected time steps $k$ for a representative run on the \textsc{ramp} terrain with the true initial state in the upper mode. Each panel shows $16$ of the $K = 256$ sampled input sequences (the $4$ highest-weighted and $12$ random ones), rolled out from the PF mean without noise.}
    \label{fig:simulations}
\end{figure}

\begin{table}[t]
    \centering
    \caption{Closed-loop comparison over $N_{\text{MC}} = 100$ runs on the \textsc{ramp} and \textsc{step} terrains (mean $\pm$ standard deviation; bold: best among the dual controllers). The time to pass is computed over successful runs, the computation time over all steps without parallel workloads.}
    \label{tab:comparison}
    \setlength{\tabcolsep}{4pt}
    \renewcommand{\arraystretch}{1.15}
    \begin{tabular}{@{}llccc@{}}
        \toprule
        & Method & \makecell{Success [\%] $\uparrow$\\{\small(collision / timeout)}} & \makecell{Time to\\pass [s] $\downarrow$} & \makecell{Comp.\ time\\per step [ms] $\downarrow$} \\
        \midrule
        \multirow{6}{*}{\rotatebox[origin=c]{90}{\textsc{ramp}}}
        & Oracle-MPPI                   & \sr{100}{0}{0} & \ms{6.9}{0.2} & \ms{9.2}{2.6} \\
        & CE-MPPI                       & \sr{6}{94}{0} & \ms{6.9}{0.2} & \ms{8.3}{0.3} \\
        & Cautious-MPPI                 & \sr{0}{0}{100} & \tbd & \ms{10.9}{0.4} \\
        \cmidrule(l){2-5}
        & DMPPI~\citep{knaup2024active} & \sr{0}{0}{100} & \tbd & \ms{16.2}{0.7} \\
        & Fisher-MPPI                   & \sr{\best{94}}{2}{4} & \ms{8.3}{1.7} & \ms{77.9}{0.9} \\
        & ED-MPPI (ours)                & \sr{91}{2}{7} & \best{\ms{8.2}{1.4}} & \best{\ms{12.7}{0.4}} \\
        \midrule
        \multirow{6}{*}{\rotatebox[origin=c]{90}{\textsc{step}}}
        & Oracle-MPPI                   & \sr{100}{0}{0} & \ms{6.9}{0.2} & \ms{8.4}{0.4} \\
        & CE-MPPI                       & \sr{5}{95}{0} & \ms{6.9}{0.2} & \ms{8.3}{0.3} \\
        & Cautious-MPPI                 & \sr{0}{0}{100} & \tbd & \ms{10.9}{0.5} \\
        \cmidrule(l){2-5}
        & DMPPI~\citep{knaup2024active} & \sr{86}{1}{13} & \ms{8.9}{2.2} & \ms{16.1}{0.6} \\
        & Fisher-MPPI                   & \sr{0}{0}{100} & \tbd & \ms{77.5}{0.9} \\
        & ED-MPPI (ours)                & \sr{\best{91}}{3}{6} & \best{\ms{8.6}{1.7}} & \best{\ms{12.6}{0.4}} \\
        \bottomrule
    \end{tabular}
\end{table}

\subsection{Results}
Figure~\ref{fig:simulations} shows a representative run on the \textsc{ramp} terrain, and Table~\ref{tab:comparison} summarizes all runs.
CE-MPPI plans from the belief mean on the center line, which appears aligned with the gap, and collides in $94\%$ of the runs.
Cautious-MPPI stalls in front of the wall in all runs, since every path through the gap collides for one mode and the direct path yields no information.
DMPPI conditions the predicted belief on the expected measurement, which, on the \textsc{ramp} terrain, shifts it towards the upper mode during a detour and thereby penalizes probing, such that DMPPI also waits in front of the wall with a bimodal belief in all runs ($k = 70$ and $k = 90$).
Fisher-MPPI probes, since the gradient of the smooth ramp indicates the informative region: it moves directly to the edge of the ramp, but remains there with an unresolved belief ($k = 30$), where the gradient is small, resolves the ambiguity at $k = 39$, and passes the gap after $\qty{8.3}{\second}$ ($k = 90$).
ED-MPPI leaves the center line while the belief is still bimodal ($k = 30$), drives onto the ramp, where the measurements remove the wrong mode ($k = 45$), and passes the gap after $\qty{8.7}{\second}$ ($k = 90$).

On the \textsc{ramp} terrain, ED-MPPI and Fisher-MPPI achieve similar success rates of $91\%$ and $94\%$, respectively.
Most failures occur for the lower mode ($84\%$ and $88\%$ success, compared to $98\%$ and $100\%$ in the upper mode), where only the particles of the other mode reach the ramp.
Neither the success rates nor the average times to pass the wall ($\qty{8.2}{\second}$ and $\qty{8.3}{\second}$) differ significantly in a paired comparison.
However, ED-MPPI is about six times faster per step than Fisher-MPPI ($\qty{12.7}{\milli\second}$ compared to $\qty{77.9}{\milli\second}$), and the Monte Carlo information term increases the computation time per step by only $\qty{1.8}{\milli\second}$ compared to Cautious-MPPI, which permits control frequencies above $\qty{70}{\hertz}$.

On the \textsc{step} terrain, the measurement gradient vanishes almost everywhere, such that Fisher-MPPI reduces to Cautious-MPPI and stalls in all runs, whereas ED-MPPI, whose information term depends only on the predicted measurements, passes the wall in $91\%$ of the runs.
DMPPI also probes on the \textsc{step} terrain ($86\%$ success), since its expected measurement stays close to the flat terrain and the predicted update removes the upper-mode particles, such that its probing depends on the shape of the measurement model rather than on the information gain.

\section{Conclusion}
We proposed ED-MPPI, a fully sampling-based explicit dual controller that scores input sequences by a belief-averaged control cost and a Monte Carlo information gain computed from the same particle rollouts, without requiring gradients or a Gaussian belief.
In a terrain-aided navigation task with a bimodal initial belief, ED-MPPI achieved a success rate similar to Fisher-information-based MPPI on smooth terrain at about one sixth of the computation time, and remained effective on piecewise-constant terrain, where the gradient-based term provides no probing incentive.

\subsection*{Generative AI Use}
Generative AI tools were used to support the literature search, mathematical derivations or proofs, the coding of the experimental implementation, and the writing and editing of the manuscript.
The authors reviewed and verified all AI-assisted content and take full responsibility for the final content of this paper.



\bibliography{ed-mppi}

\end{document}